\documentclass[a4paper,10pt]{article}
\usepackage{graphicx}
\usepackage{amssymb}
\usepackage{bm}
 \hoffset \voffset
\begin{document}

\title{\textbf{Fisher information as a performance metric for locally optimum processing}}

\author{Fabing {\sc Duan}, Fran\c{c}ois~Chapeau-Blondeau and Derek {\sc Abbott}\\
Institute of Complexity Science, Department of Automation
Engineering, \\Qingdao University, Qingdao 266071, PR China\\
Laboratoire d'Ing\'enierie des Syst\`emes Automatis\'es (LISA), \\Universit\'e d'Angers, 49000 Angers, France.
\\
Centre for Biomedical Engineering (CBME) and School of Electrical
\\\& Electronic Engineering, The University of Adelaide,\\
Adelaide, SA 5005, Australia}

\maketitle


\begin{abstract}
For a known weak signal in additive white noise, the asymptotic performance
of a locally optimum processor (LOP) is shown to be given by the Fisher information (FI)
of a standardized even probability density function (PDF) of noise in three cases:
(i) the maximum signal-to-noise ratio (SNR) gain for a periodic signal;
(ii) the optimal asymptotic relative efficiency (ARE) for signal detection;
(iii) the best cross-correlation gain (CG) for signal transmission.
The minimal FI is unity, corresponding to a Gaussian PDF, whereas the FI is
certainly larger than unity for any non-Gaussian PDFs. In the sense of a realizable LOP,
it is found that the dichotomous noise PDF possesses an infinite FI for known weak
signals perfectly processed by the corresponding LOP.
The significance of FI lies in that it provides a upper bound for the performance of locally optimum processing.
\end{abstract}


\section{Introduction}
Fisher information (FI) is a measure of the minimum error in estimating an unknown
parameter of a distribution, and its importance is related to the Cram\'{e}r-Rao inequality
for unbiased estimators \cite{Cover91,Kay98}. By introducing a location parameter,
the de Bruijn's identity indicates that the fundamental quantity of FI is affiliated
with the differential entropy of the minimum descriptive complexity of a random variable
\cite{Cover91}. Furthermore, in known weak signal detection, a locally optimum detector (LOD),
as an alternative to the Neyman-Pearson detector, has favorable properties for
small signal-to-noise ratios (SNRs) \cite{Capon61}. With sufficiently large
observed data and using the central limit theorem, it is demonstrated that the
LOD is asymptotically optimum and its asymptotic efficiency is upper bounded by
the FI of the distribution \cite{Kay98}--\cite{Kassam88}.

However, the fundamental nature of FI is not adequately recognized for
processing known weak signals. To extend the heuristic studies of \cite{Cover91}--\cite{Zozor03},
in this paper, we will theoretically demonstrate that, for a known weak signal buried
in additive white noise, the performance of a locally optimum processor (LOP) is completely
determined by the FI of a standardized even probability density function (PDF) of noise.
We show this for three signal processing case studies: (i) the maximum SNR gain for a periodic signal;
(ii) the asymptotic relative efficiency (ARE) of a LOD for signal detection;
(iii) the best cross-correlation gain (CG) for an aperidoic (random) signal transmission.
Moreover, for estimating an unknown parameter of a weak signal, the minimum mean square error
of the unbiased estimator can be reduced to a straightforward form expressed by the FI
of the distribution. The physical significance of FI, resulting from the reciprocal of
FI delimiting the minimum mean square error of unbiased estimators, provides a upper
bound of the performance for locally optimum processing. It is well known that a
standardized Gaussian PDF has minimum FI of unity \cite{Kay98}. As a consequence,
for any non-Gaussian noise, it is always possible to achieve the performance (SNR gain,
ARE or CG) of a LOP larger than unity for the three considered situations.
In the sense of a realizable LOP, an example of a Gaussian mixture noise PDF
is investigated \cite{Chapeau09}. It is found that arbitrarily large FI can be
achieved by the corresponding LOP, and even when the noise is dichotomous noise associated with infinite FI.

\section{Fisher information measures performances of locally optimum processing}
Since the known weak signal might be periodic or aperiodic,
three signal processing cases are illustrated for exploring the significance of FI in locally optimum processing.

\subsection{SNR gain for periodic signal processing}
First, consider a static processor with its output
\begin{equation}\label{system}
    y(t)=g[x(t)],
\end{equation}
where the nonlinearity $g$ is odd \cite{Zozor03} and the input is a signal-plus-noise mixture
$x(t)=s(t)+z(t)$. The component $s(t)$ is a known weak periodic signal with a maximal amplitude $A$
($0\leq|s(t)|\leq A$) and period $T$. A zero-mean white noise $z(t)$, independent of $s(t)$,
has an even (symmetric) PDF $f_z$ \cite{Lu81,Zozor03} and the root-mean-square (RMS)
amplitude $\sigma_z$ (if it exists, or it is a scale parameter.). A family of even
(symmetric) PDFs is frequently encountered in practical signal processing tasks
\cite{Kay98,Lu81,Kassam88,Zozor03}. In the case of $A\!\rightarrow \!0$, we have
a Taylor expansion around $z$ at a fixed time $t$ as
\begin{equation}\label{Taylor}
y(t)=g[z+s(t)]\approx g(z)+s(t)g'(z),
\end{equation}
where we assume the derivative $g'(z)=dg(z)/dz$ exists for almost all $x$ (similarly hereinafter)
\cite{Kay98,Kassam88} . Thus, we have
\begin{eqnarray}
{\rm E}[y(t)]&\approx& {\rm E}[g(z)]+s(t){\rm E}[g'(z)]=s(t){\rm E}[g'(z)], \label{expectation}\\
{\rm var}[y(t)]&=& {\rm E}[ y^2(t)]-{\rm E}[y(t)]^2 \approx {\rm E}[g^2(z)],\label{variance}
\end{eqnarray}
where ${\rm E}[\cdots]=\int_{-\infty}^{\infty}\cdots f_z(z)dz$.
Here, for an even PDF $f_z$ and the odd function $g$, ${\rm E}[g(z)]=0$, ${\rm E}[g(z)g'(z)]=0$ \cite{Zozor03}.
The higher order infinitesimal $s^2(t)\{{\rm E}[g'^2(z)]-{\rm E}^2[g'(z)]\}$ is neglected \cite{Zozor03},
resulting in Eq.~(\ref{variance}).

The input SNR at $x(t)$ can be defined as the power contained in the spectral line
$1/T$ divided by the power contained in the noise background in a small frequency bin
$\Delta B$ around $1/T$ \cite{Chapeau97b}, this is
\begin{equation}\label{inSNR}
    R_{\rm in}=\frac{|\langle s(t) \exp[-i 2\pi t/T]\rangle|^2}{\sigma_z^2 \Delta B \Delta t},
\end{equation}
with $\Delta t$ indicating the time resolution or the sampling period in a
discrete-time implementation and the temporal average defined as
$\langle \cdots \rangle=\frac{1}{T} \int_0^{T} \cdots dt$ \cite{Chapeau97b}.
Since $s(t)$ is periodic, $y(t)$ is in general be a cyclostationary random signal
with period $T$ \cite{Chapeau97b}. Similarly, the output SNR at $y(t)$ is given by
\begin{equation}\label{outSNR}
    R_{\rm out}=\frac{|\langle{\rm E}[y(t)]\exp[-i 2\pi t/T]\rangle|^2}{\langle{\rm var}[y(t)]\rangle \Delta B \Delta t},
\end{equation}
with nonstationary expectation ${\rm E}[y(t)]$ and nonstationary variance
${\rm var}[y(t)]$ \cite{Chapeau97b}. Here, we assume the sampling time
$\Delta t \ll T$ and observe the output $y(t)$ for a sufficiently large time interval
of $NT$ ($N\gg 1$) \cite{Chapeau97b}. Substituting Eqs.~(\ref{expectation}) and~(\ref{variance}) into Eq.~(\ref{outSNR})
and noting Eq.~(\ref{inSNR}), we have
\begin{eqnarray*}\label{approSNR}
R_{\rm out}\!\approx\!  \frac{|\langle s(t) \exp[-i 2\pi t/T_s]\rangle|^2}{\Delta B \Delta t}
\frac{{\rm E}^2[g'(z)]}{{\rm E}[g^2(z)]}\!=\!R_{\rm in}  \;\sigma_z^2 \frac{{\rm E}^2[g'(z)]}{{\rm E}[g^2(z)]}.
\end{eqnarray*}
Thus, the output-input SNR gain $G_R$ of Eq.~(\ref{system}) is
\begin{eqnarray}\label{gain}
G_R=\frac{R_{\rm out}}{R_{\rm in}}\approx  \sigma_z^2 \frac{{\rm E}^2[g'(z)]}{{\rm E}[g^2(z)]}
\leq \sigma_z^2 {\rm E}\left[\frac{f'^2_z(z)}{f_z^2(z)}\right]=\sigma_z^2 I(f_z) ,
\end{eqnarray}
with the equality occurring as $g$ becomes a LOP, viz.
\begin{eqnarray}\label{LOP}
g(z)=C f'_z(z)/f_z(z) \triangleq g_{\rm opt}(z),
\end{eqnarray}
by the Schwarz inequality for a constant $C$ and $f'_z(z)=df_z(z)/dz$ \cite{Kay98,Kassam88,Zozor03}.
It is noted that the LOP $g_{\rm opt}$ of Eq.~(\ref{LOP}) is odd and accords with the above assumption.
More interestingly, the expectation ${\rm E}\left[f'^2_z(z)/f_z^2(z)\right]$ in Eq.~(\ref{gain}) is
just the FI $I(f_z)$ of the even noise PDF $f_z$ \cite{Kay98,Kassam88}.
Furthermore, for an even standardized PDF $f_{z_0}$ with zero mean and unity variance
$\sigma_{z_0}^2=1$, the scaled noise $z(t)=\sigma_z z_0(t)$ has its PDF $f_z(z)=f_{z_0}(z/\sigma_z)/\sigma_z$.
Since the FI satisfies $I(f_z)=I(f_{z_0})/\sigma_z^{2}$ \cite{Cover91,Stam59}, the output-input SNR gain
$G_R$ of Eq.~(\ref{gain} is upper bounded by the FI $I(f_{z_0})$ of a standardized PDF $f_{z_0}(z_0)$, viz.
\begin{eqnarray}\label{Fisher}
G_R\leq I(f_{z_0}),
\end{eqnarray}
with equality achieved when $g$ takes the LOP $g_{\rm opt}$ of Eq.~(\ref{LOP}).

\subsection{Performance of a LOD for signal detection}
Secondly, we observe a data vector $X=\{x_1,x_2,\cdots,x_N\}$ composed of
$N$ observation components $x_n$, which might be the white noise $z_n$ or the mixture of
a signal $s_n$ plus white noise $z_n$ \cite{Kay98,Kassam88}. Consider a generalized correlated detector
\begin{equation}\label{GCD}
T_{GC}(X)=\sum_{n=1}^Ng(x_n) s_n > \gamma,
\end{equation}
with a memoryless nonlinearity $g$ and the decision threshold $\gamma$ for the
hypotheses $H_1: x_n=s_n+z_n$, otherwise the
hypotheses $H_0: x_n=z_n$ \cite{Kassam88}. Also, we assume that the odd function $g$ has zero mean under the even PDF $f_z$,
i.e.~${\rm E}[g(x)]=0$, and there exists a finite bound $A$ such that $0\leq|s_n| \leq A$.
In the asymptotic case of $A \!\rightarrow\! 0$ and $N\!\!\rightarrow\!\! \infty$, the test statistic
$T_{GC}$, according to the central limit theorem, converges to a Gaussian distribution with mean
${\rm E}[T_{GC}|H_0]=0$ and variance ${\rm var}[T_{GC}|H_0]\approx{\rm E}[g^2(x)]\sum_{n=1}^Ns_n^2$
under the null hypotheses $H_0$. Using Eqs.~(\ref{expectation}) and~(\ref{variance}), $T_{GC}$ is
asymptotically Gaussian with mean ${\rm E}[T_{GC}|H_1]\approx{\rm E}[g'(x)]\sum_{n=1}^Ns_n^2$
and variance ${\rm var}[T_{GC}|H_1]={\rm var}[T_{GC}|H_0]$ under the hypothesis
$H_1$ \cite{Kay98,Kassam88}. Then, given a false alarm probability $P_{FA}$,
the detection probability of the detector of Eq.~(\ref{GCD}) is expressed as
\begin{eqnarray}\label{eq:Pd}
P_D=Q\bigl[Q^{-1}(P_{FA})-\sqrt{D}\;\bigr],
\end{eqnarray}
with $Q(x)=\int_x^\infty \exp[-t^2/2]/\sqrt{2\pi}\;dt$ and its inverse function $Q^{-1}(x)$ \cite{Kay98}.
It is seen that $P_D$ is a monotonic increasing function of the deflection coefficient $D$ \cite{Kay98,Kassam88} given by
\begin{eqnarray}\label{eq:deflection}
D&=&\frac{({\rm E}[d|H_0]-{\rm E}[d|H_1])^2}{{\rm var}[d|H_0]}
\approx \frac{{\rm E}^2[g'(x)]}{{\rm E}[g^2(x)]}\sum_{n=1}^Ns_n^2\nonumber \\
&\leq & I(f_z) \sum_{n=1}^Ns_n^2 =I(f_{z_0})\sum_{n=1}^N s_n^2/\sigma_z^2,
\end{eqnarray}
with equality being achieved when $g(x)= g_{\rm opt}(x)$ of Eq.~(\ref{LOP}) \cite{Kay98,Kassam88}.
This result indicates that the asymptotic optimum detector is the LOD established by the Taylor
expansion of the likelihood ratio test statistic $\ln[ \prod_{n=1}^N f_z(x_n-s_n)/\prod_{n=1}^N f_z(x_n)]\approx \sum_{n=1}^N g_{\rm opt}(x_n) s_n$
($C=-1$) in terms of the generalized Neyman-Pearson lemma \cite{Kay98,Kassam88}.
Based on the Bayesian criterion, two hypotheses $H_0$ and $H_1$ are endowed with prior probabilities
$P_0$ and $P_1=1-P_0$. Similarly, for the weak signal $s_n$ and the sufficiently large $N$, the test statistic
$T_{GC}$ in Eq.~(\ref{GCD}) has Gaussian distribution and its performance is evaluated by the error probability \cite{Kay98}
\begin{eqnarray}\label{Bayesian}
P_{\rm e}= P_0 Q\left(\!\frac{\ln(\frac{P_0}{P_1})}{\sqrt{D}}+\frac{\sqrt{D}}{2}\right)
+P_1Q\!\left(\!\frac{\sqrt{D}}{2}-\frac{\ln(\frac{P_0}{P_1})}{\sqrt{D}}\!\right),
\end{eqnarray}
which is also a monotonically decreasing function of $D$ and has a minimum as
$D=I(f_{z_0})\sum_{n=1}^N s_n^2/\sigma_z^2$ of Eq.~(\ref{eq:deflection}) for
$g(x)=g_{\rm opt}(x)$ \cite{Kay98,Kassam88}. Interestingly, with
$D_G=\sum_{n=1}^N s_n^2/\sigma_z^2$ (called the signal energy-to-noise ratio of the data vector
$X$ \cite{Kay98}) achieved by a matched filter as a benchmark \cite{Lu81,Kassam88},
the asymptotic relative efficiency (ARE)
\begin{eqnarray}\label{ARPF}
\xi_d=D/D_G\leq I(f_{z_0}),
\end{eqnarray}
provides an asymptotic performance improvement of a detector of Eq.~(\ref{GCD})
over the linear matched filter \cite{Lu81} when both detectors operate in the same noise
environment \cite{Kay98,Lu81,Kassam88}. The equality of Eq.~(\ref{ARPF}) is
achieved as $g(x)= g_{\rm opt}(x)$ \cite{Kassam88}.

\subsection{Correlation for an aperidoic (random) signal transmission}
Thirdly, we transmit a known weak aperiodic signal $s(t)$ through the nonlinearity $g$ of
Eq.~(\ref{system}) \cite{Collins95}. Here, the signal $s(t)$ is with the average signal variance
${\rm E}[s^2(t)]=\sigma_s^2\ll \sigma_z^2$, the zero mean ${\rm E}[s(t)]=0$ and the upper bound A
($0\leq |s(t)|\leq A$). For example, $s(t)$ can be a sample according to a uniformly distributed
random signal equally taking values from a bounded interval. The input cross-correlation coefficient of
$s(t)$ and $x(t)=s(t)+z(t)$ is defined as \cite{Kay98,Collins95}
\begin{eqnarray}
\rho_{s,x}=\frac{{\rm E}[s(t)x(t)]}{\sqrt{{\rm E}[s^2(t)]}\sqrt{{\rm E}[x^2(t)]}}=
\frac{\frac{\sigma_s}{\sigma_z}}{\sqrt{\frac{\sigma_s^2}{\sigma_z^2}+1}}\approx \frac{\sigma_s}{\sigma_z}.
\end{eqnarray}
Using Eqs.~(\ref{Taylor})--(\ref{variance}), the output cross-correlation coefficient of $s(t)$ and $y(t)$ is given by
\begin{eqnarray}
\rho_{s,y}=\frac{{\rm E}[s(t)y(t)]}{\sigma_s \sqrt{{\rm var}[y(t)]}}\approx\frac{\sigma_s {\rm E}[g'(z)]}{\sqrt{{\rm E}[g^2(z)]}}.
\end{eqnarray}
Then, the cross-correlation gain (CG) $G_{\rho}$ is given by
\begin{eqnarray}\label{cg}
G_{\rho}=\frac{\rho_{s,y}}{\rho_{s,x}}\approx\sigma_z \frac{{\rm E}[g'(z)]}{\sqrt{{\rm E}[g^2(z)]}}\leq \sqrt{I(f_{z_0})},
\end{eqnarray}
which has its maximal value as $g(z)=g_{\rm opt}(z)$ of Eq.~(\ref{LOP}).

\subsection{Estimating an unknown parameter of a weak signal}
Finally, for the $N$ observation components $x_n= s_n(\theta)+z_n$,
we assume the signal $s_n(\theta)$ are with an unknown parameter $\theta$.
As the upper bound $A\rightarrow 0$ ($0\leq |s_n|\leq A$), the Cram\'{e}r-Rao
inequality indicates that the mean squared error of any unbiased estimator of the
parameter $\theta$ is lower bounded by the reciprocal of the FI \cite{Cover91,Kay98} given by
\begin{eqnarray}\label{FIest}
I(\theta)&=&\sum_{n=1}^N {\rm E}\left[\left(\frac{\partial \ln f_z(x_n-s_n)}{\partial \theta}\right)^2\right] \nonumber \\
&\approx& \sum_{n=1}^N {\rm E}\left[\left(\frac{df_z(z_n)/dz}{f_z(z_n)}\Bigr|_{z_n=x_n-s_n} \bigl(-\frac{\partial s_n}{\partial \theta}\bigr)\right)^2\right]\nonumber \\
&=&I(f_z) \sum_{n=1}^N \Bigl(\frac{\partial s_n}{\partial \theta}\Bigr)^2= \frac{I(f_{z_0})}{\sigma_z^2} \sum_{n=1}^N \Bigl(\frac{\partial s_n}{\partial \theta}\Bigr)^2,
\end{eqnarray}
which indicates that the minimum mean square error of any unbiased estimator is
also determined by the FI $I(f_{z_0})$ of distribution with a location shift,
as $\sum_{n=1}^N \bigl(\frac{\partial s_n}{\partial \theta}\bigr)^2/\sigma_z^2$ is fixed.

Therefore, just as the FI represents the lower bound of the mean squared error of
any unbiased estimator in signal estimation, the physical significance of the
FI $I(f_{z_0})$ is that it provides a upper bound of the performance for
locally optimum processing for the three considered problems.

\section{Extreme value of FI of a standardized PDF}
Some interesting questions arise: Which type of noise PDF has a minimal or maximal
FI $I(f_{z_0})$, and how large is the extreme value of $I(f_{z_0})$?
Does the corresponding LOP in Eq.~(\ref{LOP}) exist for the noise PDF with extreme
$I(f_{z_0})$? These questions will be investigated as follows.

\subsection{Minimal Fisher information of a standardized PDF}
For a standardized even PDF $f_{z_0}$, we have
\begin{eqnarray}
I(f_{z_0})={\rm E}\left[\frac{f'^2_{z_0}(z_0)}{f_{z_0}^2(z_0)} \right]
{\rm E}\left[z_0^2 \right] \geq {\rm E}\left[\frac{f'_{z_0}(z_0)}{f_{z_0}(z_0)}\; z_0 \right]^2\!=\!1,
\end{eqnarray}
with ${\rm E}\left[z_0^2\right]\!\!=\!\!\sigma_{z_0}^2\!\!=\!\!1$ and the equality occurring
if $f'_{z_0}(z_0)/f_{z_0}(z_0)=c z_0$ for a constant $c\neq 0$.
Then, $f_{z_0}(z_0)=\exp\left[k+cz_0^2/2\right]$ \cite{Kay98}. In order to be a PDF,
$c<0$ and $\exp(k)$ is the normalized constant \cite{Kay98}. This is a standardized Gaussian
PDF $f_{z_0}(z_0)=\exp(-z_0^2/2)/\sqrt{2\pi}$. Contrarily, any standardized non-Gaussian PDF
$f_{z_0}$ has the FI $I(f_{z_0})>1$, which indicates that the performance
(SNR gain, ARE or CG) is certainly larger than unity via a LOP of Eq.~(\ref{LOP})
for processing a known weak signal \cite{Kay98,Kassam88,Chapeau97b}.

\subsection{Maximal Fisher information of a standardized PDF}
A standardized generalized Gaussian noise PDF \cite{Kay98}
\begin{eqnarray}\label{ggauss}
f_{z_0}(z_0)=c_1(\beta)\exp\left[-c_2(\beta)\left|z_0\right|^{\frac{2}{1+\beta}}\right],
\end{eqnarray}
with $c_1(\beta)\!\!=\!\!\frac{1}{(1+\beta)}\frac{\Gamma^{\frac{1}{2}}\left(\frac{3}{2}(1+\beta)\right)
}{\Gamma^{\frac{3}{2}}\left(\frac{1}{2}(1+\beta)\right)}$ and $c_2(\beta)\!\!=\!\!\left|\frac{\Gamma\left(\frac{3}{2}(1+\beta)\right)
}{\Gamma\left(\frac{1}{2}(1+\beta)\right)}\right|^{\frac{1}{1+\beta}}$.
The FI $I(f_{z_0})$ of Eq.~(\ref{ggauss}) becomes \cite{Kay98}
\begin{eqnarray}
I(f_{z_0})=\frac{4}{(1+\beta)^2}\frac{\Gamma\left[\frac{3}{2}(1+\beta)\right]
\Gamma\left(\frac{3}{2}-\frac{1}{2}\beta\right)}{\Gamma^2\left[\frac{1}{2}(1+\beta)\right]},
\end{eqnarray}
with the corresponding normalized LOP \cite{Kay98}
\begin{equation}\label{NLOP}
g_{\rm opt}(x)=|x|^{(1-\beta)/(1+\beta)}{\rm sign}(x),
\end{equation}
where ${\rm sign}(\cdot)$ is the sign or signum function.
The curve of $I(f_{z_0})$ versus $\beta$ (cf.~Fig.~10.10 of Ref.~\cite{Kay98})
clearly indicates that, for $\beta=0$, $I(f_{z_0})=1$ is the minimum
corresponding to the standardized Gaussian PDF. It is also noted that, as
$\beta\rightarrow 3$ or $-1$, $I(f_{z_0})\rightarrow +\infty$.
Is the maximal $I(f_{z_0})$ infinite for $\beta=3$ and $\beta=-1$ or not,
and is the corresponding LOP simply implemented? The answer is negative,
because the LOP of Eq.~(\ref{NLOP}) is not realizable as $g_{\rm opt}(x)=\pm \infty$ for $|x|>1$
and $\beta=-1$. When $\beta>1$, the LOP of Eq.~(\ref{NLOP}) has a singularity at $x=0$.
In this sense, an arbitrary large FI cannot be reached for the generalized Gaussian noise given in Eq.~(\ref{ggauss}).

\begin{figure}
\begin{center}
\vspace{-3mm}
\includegraphics[scale=0.27]{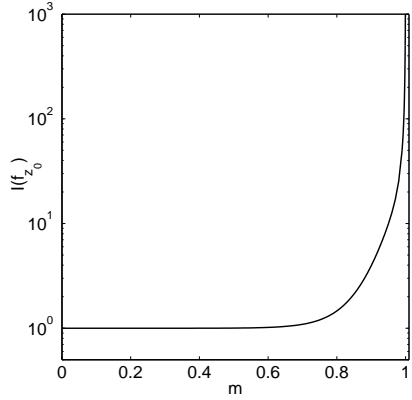}
\caption{FI $I(f_{z_0})$ versus parameter $m$ of the Gaussian mixture noise PDF of Eq.~(\ref{Gaussmix}).}  \label{fig:one}
\vspace{-6mm}
\end{center}
\end{figure}

Next, we consider Gaussian mixture noise $z(t)$ with its PDF
\begin{eqnarray}\label{Gaussmix0}
f_z(z)\!\!=\!\!\frac{1}{2\sqrt{2\pi \epsilon^2}}\!\left[\exp\Bigl(\frac{-(z-\mu)^2}{2\epsilon^2}\Bigr)
\!\!+\!\exp\Bigl(\frac{-(z+\mu)^2}{2\epsilon^2}\Bigr)\!\right]\!,
\end{eqnarray}
with variance $\sigma_z^2=\mu^2+\epsilon^2$ and parameters $\mu, \epsilon \geq 0$. Note that
Eq.~(\ref{Gaussmix0}) has another expression \cite{Zozor03} as
\begin{eqnarray}\label{GaussmixTan}
f_{z}(z)=\exp\left[-y(z)\right]/\sqrt{2\pi\epsilon^2},
\end{eqnarray}
with $y(z)=\frac{z^2+\mu^2}{2\epsilon^2}-\ln \bigl[\cosh\bigl(\frac{\mu z}{\epsilon^2}\bigr)\bigr]$.
Based on Eq.~(\ref{GaussmixTan}), the corresponding normalized LOP can be expressed as
\begin{eqnarray}\label{processorGM}
g_{\rm opt}(x)=x-\mu \tanh\left(\frac{\mu x}{\epsilon^2}\right).
\end{eqnarray}
For $0\leq m \leq 1$, assume $\mu=m \sigma_z$ and $\epsilon^2=(1-m^2)\sigma_z^2$,
Eq.~(\ref{GaussmixTan}) becomes a standardized Gaussian mixture PDF \cite{Chapeau09}
\begin{eqnarray}\label{Gaussmix}
f_{z_0}(z_0)=\exp[-y(z_0)]/\sqrt{2\pi(1-m^2)},
\end{eqnarray}
with $y(z_0)=\frac{z_0^2+m^2}{2(1-m^2)}-\ln \bigl[\cosh\bigl(\frac{m z_0}{1-m^2}\bigr)\bigr]$.
The function of FI $I(f_{z_0})$ versus $m$ is shown in Fig.~\ref{fig:one}, and $I(f_{z_0})$ can
be calculated as (no explicit expression exists)
\begin{eqnarray}\label{mgFisher}
I(f_{z_0})={\rm E}\left\{\Bigl[\frac{z_0}{1-m^2}-\frac{m}{1-m^2}\tanh\bigl(\frac{mz_0}{1-m^2}\bigr)\Bigr]^2\right\}.
\end{eqnarray}

Interestingly, Fig.~\ref{fig:one} shows that, as $m=0$, Eq.~(\ref{Gaussmix})
is the standardized Gaussian PDF with the FI $I(f_{z_0})=1$. While, $I(f_{z_0})\rightarrow +\infty$ as $m\rightarrow 1$.
In Eq.~(\ref{Gaussmix0}), for $m\rightarrow 1$, $\mu=m\sigma_z\rightarrow \sigma_z$ and
$\epsilon^2=(1-m^2)\sigma_z^2 \rightarrow 0$, the term
$\lim_{m\rightarrow 1}\frac{1}{\sqrt{2\pi\epsilon^2}}\exp\bigl[-\frac{(z-\mu)^2}{2\epsilon^2}\bigr]=\delta(z)$
and $\delta(\cdot)$ is the Dirac delta function. Then, Eq.~(\ref{Gaussmix0}) becomes
\begin{eqnarray}\label{Gaussmix2}
f_z(z)=\frac{1}{2}[\delta(z-\sigma_z)+\delta(z+\sigma_z)],
\end{eqnarray}
which represents the PDF of dichotomous noise $z(t)$. We also note
$\lim_{\epsilon\rightarrow 0,\mu\rightarrow \sigma_z} \tanh \bigl(\frac{\mu x}{\epsilon^2}\bigr)={\rm sign}(x)$ ($\mu>0$)
in Eq.~(\ref{processorGM}), and the normalized LOP for dichotomous noise $z(t)$ is
\begin{eqnarray}\label{LOPopt}
g_{\rm opt}(x)=x-\sigma_z {\rm sign}(x).
\end{eqnarray}
Here, the normalized LOP $g_{\rm opt}$ is not continuous at $x=0$,
but the above analysis is valid for processing a known weak signal in dichotomous noise.
This point is like the LOP $g_{\rm opt}(x)={\rm sign}(x)$ for Laplacian noise \cite{Kay98}.

When $z(t)$ randomly takes two levels $-\sigma_z$ and $+\sigma_z$ and $s(t)$ is weak compared
with $z(t)$ ($\sigma_z>|s(t)|$), the signs of input data $x(t)=s(t)+z(t)$ always take
the sign of $z(t)$, i.e.~${\rm sign}(x)={\rm sign}(z)$ in Eq.~(\ref{LOPopt}). Therefore,
the LOP of Eq.~(\ref{LOPopt}) at a fixed time $t$ can be solved as
$g[x(t)]=x(t)-\sigma_z \; {\rm sign}[x(t)]=s(t)+z(t)-\sigma_z {\rm sign}[z(t)]=s(t)$.
Moreover, Refs.~\cite{Chapeau97b,DeWeese95} have pointed out that there exists a scheme
allowing a perfect recovery of $s(t)$ corrupted by dichotomous noise $z(t)$ with the PDF of
Eq.~(\ref{Gaussmix2}). However, a practical difficulty in Eq.~(\ref{NLOP}) is that the
RMS $\sigma_z$ needs to be known. The above analysis indicates that the LOP of Eq.~(\ref{LOPopt})
can recover the weak signal $s(t)$ perfectly as $\sigma_z>|s(t)|$. Thus, according to the optimum
performance of the LOP of Eq.~(\ref{Fisher}) (Eq.~(\ref{ARPF}) or Eq.~(\ref{cg})), the FI $I(f_{z_0})=\infty$
contained in the type of PDF of Eq.~(\ref{Gaussmix2}), as shown in Fig.~\ref{fig:one}.
Using Eq.~(\ref{GaussmixTan}), the FI $I(f_{z_0})$ of Eq.~(\ref{mgFisher}) can be computed as
\begin{eqnarray}
I(f_{z_0})&=&{\rm E}\left[\Bigl(\frac{dy(z_0)}{dz_0}\Bigr)^2\right]={\rm E}\left[\frac{d^2y(z_0)}{dz_0^2}\right]
\nonumber \\&=&\lim\limits_{ m\rightarrow 1} \int_{-\infty}^{\infty}\!\!\! \frac{\bigl[\! 1\!-\!m^2\!-\!m^2{\rm sech}^2\!\bigl(\frac{m z_0}{1\!-\!m^2}\bigr)\bigr]}{(1-m^2)^2}\;\frac{\exp[-y(z_0)]}{\sqrt{2\pi (1-m^2)}} dz_0\!=\!\infty,
\end{eqnarray}
where $\lim_{m\rightarrow 1}\Bigl[m^2{\rm sech}^2\bigl(\frac{m z_0}{1-m^2}\bigr)\Bigr]=0$,
the numerator is the infinitesimal $\textsl{O}(1-m^2)$ and the denominator is a higher-order
infinitesimal $\textsl{O}((1-m^2)^2)$ in the integral.

\section{Conclusions}
In this paper, for a known weak signal in additive white noise,
it was theoretically demonstrated that the optimum performance of a LOP can
be completely determined by the FI of the corresponding standardized even noise PDF,
as illustrated by three signal processing case stuides: (i) the maximum SNR gain for a periodic signal;
(ii) the optimal ARE for signal detection; (iii) the best CG for signal transmission.
Thus, our study of the performance of a LOP focused on the measure of the FI of a standardized
noise PDF. It is well known that the minimal FI is unity for a standardized Gaussian noise PDF,
and the matched filter is the corresponding optimal processor.
While, for any non-Gaussian noise, the FI and hence the optimum performance of the LOP is
certainly larger than unity. Illustratively, we observed that the generalized Gaussian noise
PDF and the Gaussian mixture noise PDF have an arbitrary large FI.
There are some types of noise PDF possessing an infinite FI, such as uniform noise and dichotomous noise.
However, we argue that only if the LOP is practically realizable, can the performance predicted by the
FI be reached in practice. In this sense, it is found that the dichotomous noise has an infinite
FI and also a simple LOP structure can be realized in practice.

Some interesting questions arise. For instance, it is known that for a weak signal already
corrupted by initial additive white noise, there is usually a LOP that yields the maximal output-input gain.
Therefore, can the method of adding an extra amount of noise \cite{Chapeau09} to the initial data improve
the performance of the updated LOP for the resulting noise PDF? This interesting topic will
invoke the stochastic resonance phenomenon \cite{Zozor03,Chapeau97b,Chapeau09,Collins95,Kay00}.
Another important question is the influence of finite observation time on the performance of locally
optimum processing \cite{Chapeau09}.


\end{document}